\documentclass[11pt, a4paper]{amsart}
\usepackage{amsmath}
\usepackage{amssymb}
\def\eq#1{(\ref{#1})}
\newtheorem{thm}{Theorem}
\newtheorem{prop}[thm]{Proposition}
\newtheorem{lemma}[thm]{Lemma}
\newtheorem{corol}[thm]{Corollary}
\newtheorem{rem}[thm]{Remark}
\newtheorem{exam}[thm]{Example}
\newtheorem{defi}[thm]{Definition}

\renewcommand{\leq}{\leqslant}
\renewcommand{\geq}{\geqslant}
\newcommand{\pf}{\noindent{\it Proof \ }}
\newcommand{\epf}{{$\quad$ \hfill $\Box$}}

\def\de{\delta}
\def\ep{\epsilon}
\def\pa{\partial}
\def\Z{\mathbb{Z}}
\def\C{\mathbb{C}}

\def\cA{\mathcal{A}}

\def\cG{\mathcal{G}}
\def\cL{\mathcal{L}}
\def\cW{\mathcal{W}}

\def\cAh{\hat{\cA}}

\def\cGh{\hat{\cG}}

\def\tQ{\tilde{Q}}

\def\tu{\tilde{u}}
\def\L{\Lambda}
\def\la{\lambda}
\def\im{\mathrm{Im}}
\def\Ker{\mathrm{Ker}}
\DeclareMathOperator{\res}{Res}
\DeclareMathOperator{\tr}{Tr}

\def\sig{(1+\L)^{-1}}

\def\cF{\mathcal{F}}

\begin{document}
\title{The Extended Bigraded Toda hierarchy}
\author{Guido Carlet}
\begin{abstract}
We generalize the Toda lattice hierarchy by considering $N+M$ dependent variables. We construct roots and logarithms of the Lax operator which are uniquely defined operators with coefficients that are $\ep$-series of differential polynomials in the dependent variables, and we use them to provide a  Lax  pair definition of the {\it extended bigraded Toda hierarchy}, generalizing \cite{ca-du-za04}. Using $R$-matrix theory we give the bihamiltonian formulation of this hierarchy and we prove the existence of a tau function for its solutions. Finally we study the dispersionless limit and its connection with a class of Frobenius manifolds on the orbit space of the extended affine Weyl groups $\tilde{W}^{(N)}(A_{N+M-1})$ of the $A$ series, defined in \cite{du-za98}.
\end{abstract}
\maketitle

%%%%%%%%
\section{Introduction}
The Toda lattice \cite{to67} hierarchy is a well-studied bihamiltonian \cite{ku85} integrable system consisting of an infinite complete set of commuting evolutionary differential-difference equations for two dependent variables $u(n)$, $v(n)$, where the independent spatial variable $n$ is discrete.

If we replace the discrete variable $n$ with a continuous variable $x= n \ep$, where $\ep$ is the step of the lattice, the usual Toda flows do not form anymore a complete family. In order to have completeness one introduces a second set of flows that commute between themselves and with the usual Toda flows. These two sets of flows define the {\it extended Toda hierarchy} which was introduced in \cite{za02, ge01, ca-du-za04}. 

In particular in \cite{ca-du-za04} it was shown how to define differential-difference operators $\log_{\pm} L$ which are logarithms of the Lax operator $L= \L + u(x) + e^{v(x)} \L^{-1}$ and use them to produce a Lax representation of the extended Toda hierarchy. 

The main motivation for considering such continuous version of the Toda hierarchy comes from the applications in 2D topological field theory and in the theory of Gromov-Witten invariants. By looking at matrix models \cite{eg-ya94} describing in the large $N$ limit the $\mathbb{C}P^1$ topological sigma model, it was first conjectured and then shown \cite{du-za04} that the extended Toda hierarchy is the hierarchy describing the Gromov-Witten invariants of $\mathbb{C}P^1$. We hope that the bigraded extended hierarchy considered here might be relevant for similar applications.

In this paper we will generalize the extended Toda hierarchy by considering a Lax operator of the form
\begin{equation}
\label{fine11}
L= \L^N + u_{N-1} \L^{N-1} + \dots + u_{-M} \L^{-M}
\end{equation}
for $N,M \geq1$. If we consider this system with a discrete independent variable $n$ then we are only able to define a single set of flows corresponding to Lax equations of the form
\begin{equation}
\label{ }
L_{t_p} = [ (L^p)_+, L ] .
\end{equation}
However here we will consider a continuous independent variable $x$. 
In this case we can introduce roots $L^{\frac1N}$, $L^{\frac1M}$ and logarithms $\log_{\pm} L$ of $L$ and we can use them to define $N+M$ sequences of commuting flows (see Definition \ref{deflax})
\begin{equation}
\label{}
\ep L_{t^{\alpha,p}} = [ A_{\alpha,p} , L ] 
\end{equation}
for $-M \leq \alpha \leq N-1$ and $p\geq0$. 
We call this hierarchy the {\it extended bigraded Toda hierarchy}. 

The definitions of the roots and of the logarithms of $L$ involve certain non-localities due to the inversion of the discrete derivative operator  $\L-1$. In our approach however these non-localities are not a problem since we work with a ring of functions which are power series in $\ep$. In this case the coefficients of the roots and of the logarithms are uniquely defined $\ep$-power series of polynomials in the basic variables $u_i$.

The bihamiltonian formulation of this hierarchy is crucial in the framework of the classification program of \cite{du-za01}. Compatible Poisson brackets on the space of operators of the form \eq{fine11} were obtained for particular values of $N$, $M$ in \cite{ku85}, \cite{bl-ma94} and in the context of $q$-deformed $\mathcal{W}$-algebras in \cite{fr-re96}, \cite{fr96} for $M=0$ and in \cite{ma-se95} for general $N$, $M$. Here we derive the form of the Poisson pencil for this hierarchy in the difference operator notation using the $R$-matrix theory developed in \cite{oe-ra89,li-pa89}. The Hamiltonians are expressed in terms of powers of roots and logarithms of the Lax operator $L$. The corresponding Hamiltonian densities, and indeed the flows of the hierarchy are normalized in such a way that a certain 'tau-symmetry' holds. This turns out to be the key element in proving the existence of a tau-function for the solutions of the hierarchy.

The classification theory of tau-symmetric bihamiltonian integrable hierarchies of Dubrovin and Zhang \cite{du-za01} considers such systems in the light of their relationship with 2D topological field theory and Gromov-Witten invariants. In this framework the main invariant associated with an integrable hierarchy, or better with its dispersionless limit, is a Frobenius manifold. The main result of such classification scheme is the fact that one can reconstruct the dispersive hierarchy from the Frobenius manifold. Different classes of Frobenius manifolds have been constructed and in most cases the Lax formulation of the corresponding dispersive hierarchy is still unknown. In \cite{du-za98} it was shown that a Frobenius structure is naturally defined on the orbit space of the extended affine Weyl group $\tilde{W}^{(k)}(R)$ associated with any irreducible reduced root system $R$ with a choice of a simple root $k$. We show that the Frobenius manifold associated to the dispersionless limit of the extended bigraded Toda hierarchy is the one defined on the orbit space of $\tilde{W}^{(N)}(A_{N+M-1})$. Hence the extended bigraded Toda hierarchy coincides with the hierarchy associated to the Frobenius manifold $M_{\tilde{W}^{(N)}(A_{N+M-1})}$ by the general scheme of \cite{du-za01}.

Plans for future work include the description of the Virasoro symmetries of the extended bigraded Toda hierarchy, generalizing \cite{du-za04}, the formulation of the Hirota quadratic equations (as was recently done for the extended Toda hierarchy in \cite{mi05}) and the Lax description of hierarchies associated to Legendre transformations of the Frobenius manifold $M_{\tilde{W}^{(N)}(A_{N+M-1})}$ (see e.g. the definition of the extended NLS hierarchy in \cite{ca-du-za04}).

%%%%%
The paper is organized as follows. 
In Section 2 we define the roots and the logarithms of the Lax operator $L$. In particular we show that their coefficients are uniquely determined $\ep$-series of differential polynomials in the dependent variables.
In Section 3 we define the extended bigraded Toda hierarchy in the Lax formulation. The bi-Hamiltonian formulation of the hierarchy is given in Section 4. We provide a pair of compatible Poisson brackets and define a set of Hamiltonians which are in involution with respect to both brackets and which are explicitly given in terms of traces of the roots and the logarithm of $L$. 
In Section 5 we prove the existence of a tau-function for the extended bigraded Toda hierarchy. 
In Section 6 first we derive formulas for the generating functions of the metrics and the Poisson brackets, then we consider the Frobenius manifold associated to the dispersionless limit of this hierarchy, for fixed $N$, $M$, and show that it coincides with that constructed on the orbit space of an extended affine Weyl group of the $A$-series.

Part of the results of this paper appeared in \cite{ca03}.

%%%%%%%%%%%%%%%%%%%%
\section{Fractional powers and logarithms}
Using the dressing of the Lax operator $L$ we define its roots and logarithms.
Then we prove a theorem on the structure of such operators, generalizing previously known results for the extended Toda hierarchy \cite{ca-du-za04}. Finally we prove some formulas for the exponential of $\log_{\pm} L$.

The Lax operator of the bigraded Toda hierarchy is a difference operator of the following form
\begin{equation}
  \label{a1}
  L= \L^N + u_{N-1} \L^{N-1} + \dots + u_{-M} \L^{-M}
\end{equation}
where $N,M \geq 1$ are two fixed positive integers. The variables $u_j$ are functions of the real variable $x$ and the shift operator acts on a function $f(x)$ by $\L f(x) = f(x + \epsilon )$.

The Lax operator \eq{a1} can be written in two different ways by ``dressing'' the shift operator
\begin{equation}
  \label{a4}
 L = P \L^N P^{-1} = Q \L^{-M} Q^{-1} 
\end{equation}
where the dressing operators $P$, $Q$ have the form
\begin{align}
  \label{a5}
  &P = \sum_{k \leq 0} p_k(x) \L^{k} \qquad p_0 =1, \\
  &Q = \sum_{k \geq 0} q_k(x) \L^{k} . \label{a5.1}
\end{align}

\begin{rem}
The coefficients $p_k$ are determined recursively in terms of the variables $u_j$ through the following formula
\begin{equation}
  \label{b1}
  (\L^N -1) p_{s} = - \sum_{\substack{-M \leq k \leq N-1 \\ k \geq s+N }} u_k (\L^k p_{s-k+N})
\end{equation}
for $s \leq -1$. Notice that one has to invert the discrete derivative operator. 

In the case of $Q$ first we do a gauge transformation to get rid of the term $u_{-M}$: 
\begin{equation}
  \label{b2}
  \tilde{L} = q_0^{-1} L q_0  = \L^{-M} + \dots
\end{equation}
that implies
\begin{equation}
  \label{b3}
  q_0 = e^{(1-\L^{-M})^{-1} \log u_{-M}} . 
\end{equation}
Then one obtains the coefficients $\tilde{q}_k$ of the dressing operator $\tilde{Q} = q_0^{-1} Q =  1+ \sum_{k\geq 1} \tilde{q}_k \L^k$ such that $\tilde{L} = \tQ \L^{-M} \tQ^{-1}$ through the recursion relation
\begin{equation}
  \label{b4}
  (\L^{-M} -1) \tilde{q}_{n} = - \sum_{\substack{-M+1 \leq k \leq N \\ k \leq n-M }} \tilde{u}_k (\L^k \tilde{q}_{n-M-k} )  
\end{equation}
for $n \geq 1$, where $\tilde{u}_k = u_k \frac{\L^k q_0}{q_0}$.
Then one recovers $q_k = \tilde{q}_k q_0$.
\end{rem}

The operators $P$ and $Q$ are defined up to the multiplication from the right by operators of the form $1 + \sum_{k \leq -1} c_k \L^k$ and $\sum_{k \geq 0} \hat{c}_k \L^k$ respectively, for $c_k \in \Ker (\L^N -1 )$ and $\hat{c}_k \in \Ker (\L^{-M}-1)$. We will assume that the coefficients of $P$ and $Q$ live in a space where such kernels are give by constant functions.

We now define the fractional powers $L^{\frac1N}$ and $L^{\frac1M}$. These are two operators of the form
\begin{equation}
  \label{a2}
  L^{\frac1N} = \L + \sum_{k\leq 0} a_k \L^k , \qquad L^{\frac1M} = \sum_{k \geq -1} b_k \L^k
\end{equation}
defined by the relations
\begin{equation}
  \label{a3}
  (L^{\frac1N} )^N = L , \qquad (L^{\frac1M} )^M = L .
\end{equation}
We stress that $L^{\frac1N}$ and $L^{\frac1M}$ are two different operators, even if $N=M$.

Of course an equivalent definition can be given in terms of the dressing operators
\begin{equation}
  \label{a6}
  L^{\frac1N} = P \L P^{-1}, \qquad L^{\frac1M} = Q \L^{-1} Q^{-1} .
\end{equation}

We also define two logarithms of the operator $L$ by the following formulas
\begin{subequations}  \label{a7}
 \begin{align}
  &\log_+ L = P N \epsilon \pa P^{-1} = N \ep \pa - N \ep P_x P^{-1} , \\
  &\log_- L = - Q M \ep \pa Q^{-1} = - M \ep \pa + M \ep Q_x Q^{-1} 
\end{align} 
\end{subequations}
where $\pa = \frac{d}{dx}$.
These are differential-difference operators of the form
\begin{align}
  \label{a8}
  &\log_+ L = N \ep \pa + 2 N \sum_{k > 0} w_{-k}(x) \L^{-k} ,\\
  &\log_- L = -M \ep \pa + 2 M \sum_{k \geq 0} w_k(x) \L^k . \label{a8.1}
\end{align}
We can combine them into
\begin{equation}
  \label{a9}
  \log L = \frac1{2N} \log_+ L + \frac1{2M} \log_- L = \sum_{k \in \Z} w_k \L^k 
\end{equation}
that is a pure difference operator since the derivatives cancel.

Notice how the ambiguity in the definition of $P$, $Q$ cancels in the definition of the logarithms and of the fractional powers of $L$. 

We would like to find explicit expressions for the operators $\log L$, $L^{\frac1N}$ and $L^{\frac1M}$ in terms of the coefficients of $L$. This involves the inversion of the discrete derivative operator $\L-1$ that appears in the recursive definition of the coefficients $a_k$, $b_k$ and $w_k$. It is not possible to find explicit formulas for such inverse if we work on the space $\cF$ of smooth functions of $x$. We can consider instead functions in $\cF[[\ep]]$, i.e. of the form
\begin{equation}
  \label{a10}
  f(x,\ep) = \sum_{k \geq 0} f_k(x) \ep^k ,
\end{equation}
and let the shift operator act on these functions as the exponential of the $x$-derivative
\begin{equation}
  \label{a11}
  \L f(x, \ep) = e^{\ep \frac{d}{dx}} f(x, \ep) = \sum_{k\geq 0} \frac{\ep^k}{k!} (\frac{d}{dx})^k f(x, \ep) .
\end{equation}
One can easily prove that 
\begin{equation}
  \label{a14}
  \im (\L^m -1) = \im (\ep \frac{d}{dx} ), \qquad \Ker (\L^m -1) = \C [[\ep]] .
\end{equation}
For a function $f \in \cF[[\ep]]$ we have an explicit inversion formula in terms of the Bernoulli numbers $B_k$ 
\begin{equation}
  \label{a12}
  (\L^m -1)^{-1} \ep \frac{d}{dx} f = \frac1m \sum_{k\geq 0} \frac{B_k}{k!} (m \ep \frac{d}{dx})^k f .
\end{equation}
The Bernoulli numbers are defined by the generating function
\begin{equation}
  \label{a13}
  \frac{\la}{e^\la -1} = \sum_{k\geq 0} \frac{B_k}{k!} \la^k .
\end{equation}

\begin{defi}
  We denote by $\cA$ the algebra of differential polynomials in $u_{N-1}, \dots, u_{-M+1}$,$(u_{-M})^{\frac1M}$, $(u_{-M})^{-\frac1M}$ and $\log u_{-M}$; $\hat{\cA} := \cA[[\ep]]$ is the differential algebra of formal power series in $\ep$ with coefficients in $\cA$.
\end{defi}

We define a gradation on $\hat{\cA}$ by 
\begin{equation}
  \label{b8}
  \deg u_k = 1-\frac{k}N, \quad \deg (u_{-M})^{\frac1M} = \frac1N + \frac1M,
\quad \deg \log u_{-M} = 0.
\end{equation}

On $\hat{\cA}$ we have also another gradation, associated with the number of derivatives, that we denote by $\deg_{\pa}$ and is defined by
\begin{equation}
  \label{b8.1}
  \deg_{\pa} \pa_x^k u_l = - k \quad \deg_{\pa} \ep =1 .
\end{equation}
The operators $\ep \frac{\pa}{\pa x}$ and $\L = e^{\ep \pa}$ act on $\hat{\cA}$ and preserve both gradations.

It turns out that the coefficients of $L^{\frac1N}$, $L^{\frac1M}$, $\log L$, unlike the coefficients of $P$, $Q$, can be expressed in terms of the coefficients of $L$, i.e. are elements of $\hat{\cA}$. Indeed we have
\begin{thm}
  The coefficients $a_k$, $b_k$ and $w_k$ in the definitions of the operators $L^{\frac1N}$, $L^{\frac1M}$ and $\log L$ are uniquely determined elements of $\hat{\cA}$.
\end{thm}
\pf
Spelling out the coefficient of $\L^{p}$ in the relation $(L^{\frac1N})^N =L$ that defines $L^{\frac1N}$ we have, for $p \leq N-1$
\begin{equation}
  \label{b9}
  (1 + \cdots + \L^{N-1}) a_{p-N+1} = u_p - \sum_{\substack{p-N+1 < l_i \leq 1 \\ l_1 + \cdots +l_N = p}}
  a_{l_1} (\L^{l_1} a_{l_2} )\cdots (\L^{l_1 + \cdots + l_{N-1} } a_{l_N}) .
\end{equation}
Since $1 + \cdots + \L^{N-1}$ is invertible on $\hat{\cA}$ we have that $a_m \in \hat{\cA}$.

Now consider the operator $L^{\frac1M}$. We define $\tilde{L} = q_0^{-1} L q_0=\sum_k \tilde{u}_k \L^k$ where $q_0$ is the leading term in the expansion \eq{a5.1} of the dressing operator $Q$. The coefficients $\tilde{u}_k$ are clearly elements of $\hat{\cA}$ since they are expressed as 
\begin{equation}
  \label{b10}
  \tu_k = \frac{\L^k q_0}{q_0} u_k
\end{equation}
and 
\begin{equation}
  \label{b11}
  \frac{\L^k q_0}{q_0}  = e^{(1-\L^{-M})^{-1} (\L^k -1) \log u_{-M}}. 
\end{equation}
Indeed from the definition of $Q$ one has $ \frac{q_0}{\L^{-M} q_0} = u_{-M} $ from which \eq{b11} follows. Moreover, since $(1- \L^{-M})^{-1} (\L^k -1) \log u_{-M} = \frac{k}M \log u_{-M} + \sum_{l\geq 1} g_l (\ep \pa)^{l-1} \frac{\ep (u_{-M})_x}{u_{-M}}$ for some constants $g_k$, we have that \eq{b11} equals
\begin{equation}
  \label{b12}
  (u_{-M})^{\frac{k}M}  \sum_{p \geq 0 } \frac1{p!} (\sum_{l\geq 1} g_l (\ep \pa)^{l-1} \frac{\ep (u_{-M})_x}{u_{-M}})^p
\end{equation}
hence $\frac{\L^k q_0}{q_0}$ is in $\hat{\cA}$. 

Defining $\tilde{L}^{\frac1M} = q_0^{-1} L^{\frac1M} q_0$ we then have $(\tilde{L}^{\frac1M})^M = \tilde{L}$ and substituting
$\tilde{L}^{\frac1M} = \L^{-1} + \sum_{k \geq 0} \tilde{b}_k \L^k$ we obtain
\begin{equation}
  \label{b13}
  ( 1+ \cdots + \L^{-M+1}) \tilde{b}_{p+M-1} = \tu_p - \sum_{\substack{-1 \leq l_i \leq p+M-2 \\ l_1 + \dots + l_M = p}} 
\tilde{b}_{l_1} \cdots (\L^{l_1 + \cdots + l_{M-1}} \tilde{b}_{l_M}) .
\end{equation}
From this equation and \eq{b10} it follows that $\tilde{b}_k \in \hat{\cA}$. Also $b_k \in \hat{\cA}$ since 
\begin{equation}
  \label{b14}
  b_k = \frac{q_0}{\L^k q_0} \tilde{b}_k .
\end{equation}

Now consider the coefficients $w_k$ for $k \leq -1$. Dressing the relation $[ \ep \pa , \L^m ] =0$ with the operator $P$ one obtains $[ \log_+ L , L^{\frac{m}N} ] = 0$ and taking the residue:
\begin{equation}
  \label{b15}
  (\L^m -1) w_{-m} = \frac12 \ep \frac{d}{dx} p_0(m) + \sum_{0<l<m} (\L^{-l} -1) (p_l (m) \L^l w_{-l}) 
\end{equation}
where $L^{\frac{m}N} = \sum_{k \leq m} p_k(m) \L^k$.
The RHS is clearly in the image of $\L^m -1$, so we can obtain $w_m$ for $m <0$ in terms of $w_k$, $m < k < 0$; hence $w_m \in \hat{\cA}$.
A priori each $w_m$ is determined up to an element of $\Ker(\L^m -1) = \C[[\ep]]$. We show that the constant term in each $w_m$, $m<0$ must be zero for \eq{a4} to hold. Infact if we put all $u_j$'s to zero in \eq{a4} we obtain $(\L^N -1) p_k =0$ hence $P_x = 0$; from \eq{a7} and \eq{a8} one immediately obtains $w_m =0$. Since the $w_m$ are differential polynomials in the $u_j$ this implies that the constant term is zero for each $m<0$.

Finally we consider the case of $\log_- L$. In this case one dresses with $\tilde{Q} = q_0^{-1} Q$ the relation $[\ep \pa , \L^m ]=0$ and obtains $[\log_- \tilde{L}, \tilde{L}^{\frac{m}M} ] =0$, where 
\begin{equation}
  \label{b16}
\log_- \tilde{L} = - \tilde{Q} M\ep \pa \tilde{Q}^{-1} = -M \ep \pa + 2 M \sum_{k >0} \tilde{w}_k \L^k  .
\end{equation}
Substituting and taking the residue we obtain, for $m>0$,
\begin{equation}
  \label{b17}
  (1-\L^{-m})\tilde{w}_m = \frac12 \ep \frac{d}{dx} \tilde{p}_0(m) - \sum_{k=1}^{m-1} (1-\L^{-k}) (\tilde{w}_k \L^k \tilde{p}_{-k}(m)) ,
\end{equation}
where $\tilde{L}^{\frac{m}M} = \sum_{k\geq -m} \tilde{p}_k (m) \L^k $. Thus the coefficients $\tilde{w}_k$ for $k>0$ are differential polynomials in the variables $\tilde{u}_k$. Moreover they are uniquely determined, as one can show using the same argument as above. Using \eq{b10} one goes back to the original variables $u_k$ and hence has that the coefficients
\begin{equation}
  \label{b18}
  w_k = \tilde{w}_k \frac{q_0}{\L^k q_0} 
\end{equation}
for $k>0$ are uniquely determined elements of $\hat{\cA}$.

Finally one has, by definition, 
\begin{equation}
  \label{b19}
  w_0 = \frac12 \ep \frac{d}{dx} \log q_0 = \frac12 (1- \L^{-M})^{-1} \ep \frac{d}{dx} \log u_{-M} \in \hat{\cA}.
\end{equation}
\epf

We have the following result on the degree of these coefficients
\begin{prop}
The coefficients of $L^{\frac1N}$, $L^{\frac1M}$ and $\log L$ defined in \eq{a2}, \eq{a8} and \eq{a8.1} have the degrees 
  \begin{equation}    
 \deg a_k = \frac{1-k}N , \quad \deg b_j = \frac1M -\frac{j}N , \quad \deg w_l = \frac{-l}N 
  \end{equation}
for $k \leq 0$, $j \geq -1$ and $l \in \Z$.
\end{prop}
Moreover, since the derivative $\frac{\pa}{\pa x}$ enters all the formulas above always multiplied by $\ep$, these coefficients are all degree $0$ homogeneous elements of $\hat{\cA}$  with respect to the grading $\deg_{\pa}$.

\pf 
One simply applies the $\deg$ operator to the explicit formulas in the previous proof. Notice that in particular 
\begin{equation}
  \label{b20}
  \deg \frac{\L^k q_0}{q_0} = k (\frac1N + \frac1M) \quad \text{and} \quad 
\deg ( \ep \frac{d}{dx} \log q_0) = 0 . 
\end{equation}
\epf

\begin{exam}
  Some examples of the coefficients $w_k$ of the difference operator $\log L$ are
    \begin{align}
      \label{b21}
      &w_{-1} = \frac{\ep}{2} (\L^N -1)^{-1} (u_{N-1})_x \\
      &w_0 = \frac{\ep}2 (1- \L^{-M} )^{-1} \frac{(u_{-M})_x}{u_{-M}} \label{b21.1}\\
    &w_1 = \frac{\ep}2 e^{ (1-\L^{-M})^{-1} (1-\L) \log u_{-M} }  (1-\L^{-M})^{-1} \cdot \\
    &\qquad \cdot\left( e^{(1-\L^{-M})^{-1} (\L^{-M+1} -1) \log u_{-M} } u_{-M+1} \right)_x 
    \end{align}
These are infinite series in $\ep$, e.g.
\begin{align}
  \label{b22}
  &w_{-1} = \frac1{2N} u_{N-1} - \frac14 \ep (u_{N-1})_x + O (\ep^2) \\
  &w_0 = \frac1{2M} \log u_{-M} + \frac{\ep}4 (\log u_{-M} )_x + \frac{\ep^2}{24}M (\log u_{-M} )_{xx} + O(\ep^3) \\
  &w_1 = \frac1{2M} \frac{u_{-M+1}}{u_{-M}} + \frac{\ep}{4M} \frac{M u_{-M} (u_{-M+1})_x - (M+1) u_{-M+1} (u_{-M})_x}{(u_{-M})^2} +O(\ep^2).
\end{align}
\end{exam}

%%%%%%%%%%%%%%%%%%%%%%%%
We have defined the logarithms of $L$ by dressing the derivative operator. However the usual power series definition of exponential, if applied e.g. to $\log_+ L$ : 
\begin{equation}
\label{ }
e^{\log_+ L} = \sum_{k\geq 0} \frac1{k!} (N \ep \pa + W_-)^k  
\end{equation}
produces a differential-difference operator of infinite order in $\ep \pa$ and in $\L$.
It turns out that we can reorganize the powers of $\ep \pa$ in a simple way by using the fact that $e^{\ep \pa}$ and $\L$ act in the same way on functions (considered as power-series in $\ep$). In other words 
\begin{lemma} \label{lemmacom}
The differential-difference operator $e^{\ep \pa} \L^{-1}$ commutes with any other operator.
\end{lemma}
Now it is easy to see that
\begin{prop}
The following equalities between differential-difference operators hold true:
\begin{equation}
\label{for1}
e^{\log_+ L} = L e^{N \ep \pa} \L^{-N} ,\qquad 
e^{\log_- L} = L e^{-M \ep \pa} \L^{M} .
\end{equation}
\end{prop}
\pf 
Since $e^{-N \ep \pa} \L^N$ commutes with $P^{-1}$ we have
\begin{equation}
\label{ }
L=P \L^{N} P^{-1} = P e^{N\ep \pa} P^{-1} e^{-N \ep \pa} \L^N = e^{\log_+ L} e^{-N \ep \pa} \L^N,
\end{equation}
and similarly for $\log_- L$.
\epf

One obtains that
\begin{equation}
\label{ }
\log_+ L = \log( L \L^{-N} e^{N \ep \pa}) = \sum_{k\geq 1} \frac{(-1)^{k+1}}{k} (L \L^{-N} e^{N \ep \pa} -1)^k .
\end{equation}

It is interesting to consider difference operators $\cW_-$ and $\cW_{>0}$ associated to the logarithms $\log_{\pm} L$. Recall that , see \eq{a8}-\eq{a8.1}
\begin{equation}
\label{for4}
\log_+ L = N \ep \pa + 2N W_- \quad \text{and} \quad 
\log_- L = -M \ep \pa + 2M w_0 +2M W_{>0} .
\end{equation}
where we denoted $W=\sum_{k} w_k \L^k$. We have the following
\begin{prop}
The formulas
\begin{equation}
\label{for2}
e^{\log_+ L} e^{-N\ep \pa}=e^{\cW_-} \qquad e^{\log_- L} e^{M \ep \pa -2M w_0} = e^{\cW_{>0}} 
\end{equation}
uniquely define two difference operators $\cW_-$ and $\cW_{>0}$. They can be expressed as series in $W_-$ and $W_{+}$ and their derivatives,  respectively. The first terms in these series are
\begin{align}
\label{}
    \cW_- &= 2NW_- + N^2 \ep (W_-)_x + \frac13  N^3 \ep^2 (W_-)_{xx} +\frac13 N^3 (W_- \ep (W_-)_x - \ep (W_-)_x W_-) + \dots  \\
    \cW_{>0}&= 2M W_{>0} - M^2 \ep (W_{>0})_x + 2M^2 [w_0,W_{>0}] + \dots
\end{align}
\end{prop}
\pf 
This follows from the form of the logarithms \eq{for4}, without using formulas \eq{for1}. The proof is a simple application of the Baker-Campbell-Hausdorff formula, which says that, given non-commuting variables $X$, $Y$ there exists a power series  $Z$ in $X$, $Y$ such that
\begin{equation}
\label{ }
e^X e^Y = e^Z .
\end{equation}
The first few terms of $Z$ are
\begin{equation}
\label{ }
Z = X+Y +\frac12 [X,Y] + \frac1{12} ([[Y,X],X]+[[X,Y],Y] ) + \dots
\end{equation}
and all higher order terms all involve nested commutators of $X$ and $Y$. 

If we apply the BCH formula to the product 
\begin{equation}
\label{ }
e^{N \ep \pa + 2N W_-} e^{-N\ep \pa} = e^{\cW_-}
\end{equation}
we see straight away that after the linear term $W_-$ all the higher terms are iterated commutators involving  $\ep \pa$ and $W_-$, hence the derivative operator is not present in $\cW_-$.

The same considerations hold for $\cW_{>0}$.
\epf

Comparing formulas \eq{for1} and \eq{for2} we obtain
\begin{prop}
\begin{align}
\label{ }
e^{\cW_-} \L^N &= L \quad \\
e^{\cW_{>0}} u_{-M} \L^{-M} &= L
\end{align}
\end{prop}
\pf
To prove the second formula one must use the fact that 
\begin{equation}
\label{ }
e^{-M \ep \pa + 2M w_0} e^{M \ep \pa} = e^{(2 (\ep \pa)^{-1} (1-\L^{-M}) w_0)} .
\end{equation}
Notice that on the LHS we have the product of two (formal) differential operators, while on the RHS the exponent is a function, since the operators act on $w_0$ alone. This formula is proved by dressing $-M\ep \pa$ with a function $p(x)$
\begin{equation}
\label{ }
-M \ep \pa + 2M w_0 = p (-M \ep \pa ) p^{-1} = - M\ep \pa + M \ep p_x p^{-1} 
\end{equation}
and then
\begin{equation}
\label{ }
p ( e^{-M\ep \pa} p^{-1}) = e^{((1-e^{-M\ep \pa}) \log p)}= e^{(2(\ep\pa)^{-1} (1- e^{-M\ep\pa}) w_0)}
\end{equation}

Then one uses the expression of $w_0$ in terms of $u_{-M}$, equation \eq{b21.1}.
\epf

From the previous formulas we have  that $\cW_-$ and $\cW_{>0}$ can be expressed as logarithms of $L \L^{-N}$ and $L \L^M \frac1{u_{-M}}$ respectively,
\begin{equation}
\label{ }
\cW_- = \log L \L^{-N} = \log (1+(L \L^{-N}-1)) = \sum_{k\geq1} \frac{(-1)^k}{k} (L \L^{-N} -1)^k = ...
\end{equation}
and similarly for $\cW_{>0}$, from which we obtain explicit expressions of this operators in terms of the variables $u_j$. Notice in particular that the coefficients of $\cW_-$ and $\cW_{>0}$ are difference polynomials in tha variables $u_j$, $(u_{-M})^{-1}$
\begin{align}
\label{}
    \cW_- &= u_{N-1} \L^{-1} + (u_{N-2} - \frac12u_{N-1} (\L^{-1} u_{N-1}) ) \L^{-2} + \dots   \\
    \cW_{>0} &= \frac{u_{-M+1}}{\L u_{-M}} \L + \left( \frac{u_{-M+2}}{\L^2 u_{-M}} -\frac12 
    \frac{u_{-M+1} \L u_{-M+1}}{(\L u_{-M}) \L^2 u_{-M}} \right) \L^2 + \dots  
\end{align}

%%%%%%%%%%%%%%%%%%%%%%%%%
\section{Lax formulation}
We define the flows of the bigraded extended Toda hierarchy using the Lax formalism.

Given any difference operator $A= \sum_k A_k \L^k$, the positive and negative projections are given by $A_+ = \sum_{k\geq0} A_k \L^k$ and $A_- = \sum_{k<0} A_k \L^k$.
\begin{defi} \label{deflax}
The {\rm extended bigraded Toda hierarchy} consists of the system of flows given, in the Lax pair formalism, by
\begin{equation}
  \label{b5}
  \ep \frac{\pa L}{\pa t^{\alpha, q}} = [ A_{\alpha,q} , L ] 
\end{equation}
for $\alpha = N-1, \dots , -M$ and $q \geq 0$. The operators $A_{\alpha ,q}$ are defined by 
\begin{subequations}
\label{b6}
\begin{align}
  &A_{\alpha,q} = \frac{\Gamma (2- \frac{\alpha}{N} )}{\Gamma(q+2 -\frac{\alpha}{N} ) } ( L^{q+1-\frac{\alpha}N })_+ \quad \text{for} \quad \alpha = N-1, \dots, 0 \\
  &A_{\alpha,q} = \frac{-\Gamma (2+\frac{\alpha}{M} )}{\Gamma(q+2 +\frac{\alpha}{M} ) } ( L^{q+1+\frac{\alpha}M })_- \quad \text{for} \quad \alpha = 0, \dots, -M+1 \\
  &A_{-M,q} = 2\frac1{q!} [ L^q (\log L - \frac12 ( \frac1M + \frac1N) c_q ) ]_+ .
\end{align}
\end{subequations}
The constants $c_q$ are defined by 
\begin{equation}
  \label{b24}
  c_q = \sum_{k=1}^q \frac1k , \quad c_0=0 . 
\end{equation}
\end{defi}

For $N=1=M$ this hierarchy coincides with the extended Toda chain hierarchy introduced in \cite{ca-du-za04}.

The compatibility of the flows follows from the Zakharov-Shabat equations:
\begin{prop} \label{propZS}
If $L$ satisfies the Lax equations then
\begin{equation}
\label{ZSeq}
\ep (A_{\alpha,p})_{t^{\beta,q}} - \ep (A_{\beta, q})_{t^{\alpha,p}} + [ A_{\alpha,p} , A_{\beta,q} ] =0
\end{equation}
for $-M \leq \alpha,\beta \leq N-1$ ,  $p,q \geq 0$. 
\end{prop}
Strictly speaking this should be interpreted in the following way: the vector fields defined by the Lax equations imply the Zakharov-Shabat equations, hence they commute so it's possible to find a simultaneous solution to the Lax equations. It's actually possible to show the stronger statement that the Zakharov-Shabat equations are equivalent to the Lax equations. 

Before giving the proof we need to introduce some notations which will be used repeatedly in the rest of the article. We define the following operators 
\begin{equation}
  \label{d1}
  B_{\alpha , p} := 
  \begin{cases}
  \frac{\Gamma ( 2- \frac{\alpha}N )}{\Gamma (p+2 - \frac{\alpha}N )}  L^{p+1-\frac{\alpha}N} &\alpha=0\dots N-1\\  
  \frac{\Gamma ( 2 + \frac{\alpha}M )}{\Gamma (p+2 + \frac{\alpha}M )}  L^{p+1+\frac{\alpha}M} &\alpha = 0\dots -M+1\\
 2\frac1{p!} [ L^p ( \log L - \frac12 ( \frac1M + \frac1N ) c_p ) ] & \alpha = -M.
  \end{cases}
\end{equation}
We can equivalently define the flows of the hierarchy by 
\begin{equation}
  \label{d2}
  \ep \frac{\pa L}{\pa t^{\alpha,p}} = [ (B_{\alpha,p})_+ , L ]= [- (B_{\alpha,p})_- , L ] .
\end{equation}

We have  the following  
\begin{lemma} \label{lemD}
We have that
\begin{align}
  \label{d6}
  & \ep (L^{\frac1N})_{t^{\alpha,p}} = [ - (B_{\alpha,p})_-, L^{\frac1N} ] \\
  & \ep (L^{\frac1M})_{t^{\alpha,p}} = [ (B_{\alpha,p})_+, L^{\frac1M} ] \label{d6i}\\
  & \ep (\log_+ L)_{t^{\alpha,p}} = [ -(B_{\alpha,p})_-, \log_+ L ] \label{d6ii}\\
  & \ep (\log_- L)_{t^{\alpha,p}} = [(B_{\alpha,p})_+ ,\log_- L ] . \label{d6iii}
\end{align}
and combining the last two equations 
\begin{equation}
  \label{d7}
  \ep (\log L)_{t^{\alpha,p}} = [ -(B_{\alpha,p})_-, \frac1{2N} \log_+ L ] +
[(B_{\alpha,p})_+ ,\frac1{2M} \log_- L ] .
\end{equation}
\end{lemma}
\pf
Equations \eq{d6} and \eq{d6i} are quite obvious. Since e.g.  \eq{d6} together with \eq{a3} implies \eq{b5} then it is clear that it gives the correct push-forward of the vector field \eq{b5} under the map $L \rightarrow L^{\frac1N}$. The same holds for \eq{d6i}.

To show \eq{d6ii} consider that $[\log_+ L , L^{\frac{m}N} ] =0$ and then derive this formula by $t^{\alpha,p}$. Using the Jacobi identity one obtains 
\begin{equation}
\label{d78}
[ \ep (\log_+ L)_{t^{\alpha,p}} + [\log_+ L , -(B_{\alpha,p})_- ], L^{\frac{m}N}] = 0 .
\end{equation} 
Then one notices that $\ep (\log_+ L)_{t^{\alpha,p}} + [\log_+ L , -(B_{\alpha,p})_- ]$ is an operator of the form $\sum_{l<0} f_l \L^l$ with $f_l$ homogeneous elements in $\cAh$ of strictly positive degree. 
Then taking the residue of \eq{d78} one obtains that all $f_l$ are zero, hence formula \eq{d6ii} is proved.

In a similar way one proves \eq{d6iii}.
\epf

Finally we can prove Proposition \ref{propZS}

\pf
One has to consider separately the cases $\alpha$ and $\beta$ are greater or smaller than zero or equal to $-M$ and use the previous Lemma to derive $A_{\alpha ,p}$ and $A_{\beta, q}$.
When both $\alpha, \beta \geq 0$ the proof of \eq{ZSeq} is simply given by
\begin{align}
\label{ }
\ep (A_{\beta, q})_{t^{\alpha,p}} &= [ (B_{\alpha,p})_+, B_{\beta, q} ]_+ =[ A_{\alpha,p},A_{\beta,q}]+[(B_{\alpha,p})_+,(B_{\beta,q})_-]_+ = \\ 
&= [ A_{\alpha,p},A_{\beta,q}] - [(B_{\beta,q})_-, B_{\alpha,p}]_+
=[ A_{\alpha,p},A_{\beta,q}] + \ep (A_{\alpha, p})_{t^{\beta,q}}
\end{align}
The other cases are done in a more involved but similar way. Special care must be taken when $\alpha$ or $\beta=-M$.
\epf

\begin{rem}
The choice of the normalization of the coefficients $A_{\alpha, q}$ comes from the requirement of the tau symmetry for the associated Hamiltonians (see Section 5). Anyway this choice is not unique since we can multiply $A_{\alpha, q}$ by a $c_{\alpha, q}= K^q K_{\alpha}$ for arbitrary constants $K$, $K_{\alpha}$ preserving the tau symmetry.
\end{rem}

\begin{rem}
  The operators $A_{\alpha,q}$ for $\alpha \not= -M$ are difference operators of bounded order with coefficients in $\hat{\cA}$. On the other hand $A_{-M, q}$ is a difference operator of unbounded order. We can however give an equivalent definition of these flows by 
  \begin{equation}
    \label{b25}
    \tilde{A}_{-M,q} = 2\frac1{q!} [ L^q (\log L - \frac12 (\frac1M + \frac1N) c_q ]_+
    - \frac1{q!} [ L^q ( \frac1M \log_- L - \frac12 (\frac1M + \frac1N) c_q ) ] .
  \end{equation}
This gives exactly the same flows $t^{-M,q}$ through equation \eq{b5} since it differs from (\ref{b6}c) by a part that commutes with $L$.
This operator is not purely difference, since it contains the derivative $\frac{d}{dx}$, but it contains a finite number of terms as one can see in the explicit form
\begin{equation}
  \label{b26}
  \tilde{A}_{-M,q} = \frac1{q!} L^q \ep \pa + \frac2{q!} [ L^q ( \sum_{k<0} w_k \L^k - \frac14 (\frac1M + \frac1N) c_q)]_+
  -\frac2{q!} [L^q ( \sum_{k \geq 0} w_k \L^k - \frac14 (\frac1M + \frac1N) c_q)]_-
\end{equation}
\end{rem}

\begin{exam}
  These are some explicit examples of the operators $A_{\alpha,q}$:
  \begin{align}
    \label{b27}
    &A_{N-1,0} = (L^{\frac1N})_+ = \L + ( (\L^{N-1} + \dots + 1)^{-1} u_{N-1}) , \\
    &A_{0,0} = L_+ = \L^N + \dots + u_0 , \\
    &A_{-M+1,0} = - (L^{\frac{1}M})_- = - e^{(1+ \cdots +\L^{-M+1})^{-1} \log u_{-M}} \L^{-1} , \\
    &\tilde{A}_{-M,0} =  \ep \pa .
  \end{align}
\end{exam}

Finally we have the following simple result on the gradation of the flows
\begin{prop}
  The components of the vector fields of the extended bigraded Toda hierarchy defined in \eq{b5}-\eq{b6} are homogeneous elements of the graded algebra $\hat{\cA}$ with degree
  \begin{equation}
    \label{b23}
    \deg \frac{\pa u_k}{\pa t^{\alpha,q} } = 
q+2+ \mu_{\alpha} - \frac{k}N 
%    \begin{cases}
%      (q+2) N -\alpha -k  -1 & \text{for $\alpha=N-1,\dots, 0$} \\
%      (q+2) N + \alpha\frac{N}M -k -1 &\text{for $\alpha=0, \dots, -M$}
%    \end{cases}
  \end{equation}
where $\mu_{\alpha}$ is defined by
\begin{equation} \label{b23.1}
\mu_{\alpha} = 
  \begin{cases}
     - \frac{\alpha}N &\alpha \geq 0 \\
     \frac{\alpha}M &\alpha \leq 0 .
  \end{cases} 
\end{equation}
\end{prop}
Clearly they are also homogeneous with respect to the degree $\deg_{\pa}$ with $\deg_{\pa} \ep \frac{\pa u_k}{\pa t^{\alpha,q}} =0$.

%%%%%%%%%%%%%%%%%%%%%%%%%%%%%%%%%%%%%
\section{The Hamiltonian formulation}
In this section we show that the extended bigraded Toda hierarchy is bi-Hamiltonian. We provide two compatible Poisson brackets and a set of Hamiltonians which are in involution with respect to both brackets. The Hamiltonians are explicitly given as traces of powers and logarithms of $L$ and satisfy a non--standard recursion relation.

First we define some usual notations. Given a difference operator $A= \sum_k a_k \L^k$, the residue of $A$ is given by
\begin{equation} \label{rr1}
  \res A = a_0 
\end{equation}
and the trace by
\begin{equation} \label{rr2}
  \tr A = \int dx \ \res A .
\end{equation}
Using the trace we define the inner product of two difference operators
\begin{equation} \label{rr3}
  <A, B> = \tr AB .
\end{equation}

The equations of the bigraded Toda hierarchy define flows on the space $\cG$ of difference operators of the form \eq{a1}. A tangent vector to this phase space is given by a difference operator of the form 
\begin{equation} \label{rr4}
\dot{L} = \dot{u}_{N-1} \L^{N-1} + \dots + \dot{u}_{-M} \L^{-M}
\end{equation}
while a $1$-form $w$ can be represented in terms of a difference operator $W$ using the inner product 
\begin{equation} \label{rr5}
w(\dot{L}) = < W, \dot{L} > . 
\end{equation}
The operator $W$ is not uniquely associated to the $1$-form $w$ but can actually be modified by adding any operator of the form $(\sum_{k<-N+1}+ \sum_{k>M}) w_k \L^k$.

Given a function $f(L)$ its differential at the point $L$ will be given by a difference operator $df$ such that 
\begin{equation}
\label{rr5.1}
\frac{d}{dt}f(L(t))|_{t=0} = <df, \frac{d}{dt}L|_{t=0}>
\end{equation}
where $L(t)$ is any path with $L(0)=L$.

A Poisson bracket of two functions $f$, $g$ of $L$ can be given as
\begin{equation} \label{rr6}
\{ f, g\} = <df, P (dg)> 
\end{equation}
where $df$, $dg$ are two difference operators that represent the differentials of the functions  $f$, $g$ as in \eq{rr5.1} and the Hamiltonian operator $P$ maps 1-forms to vectors.

\begin{prop}
The following formulas
\begin{align} \label{e7}
	  P_1 (X) = &[ X_+ , L ]_{\leq 0} -[X_- , L ]_{>0} \\
	  P_2 (X) = &\frac12[ L,(LX+XL)_-] - \frac12L[L,X]_{\leq0}- \frac12 [L,X]_{\leq0} L    \label{e7b}  \\
	  &+\frac12 [L, ((\L^{N}+1)(\L^{N}-1)^{-1}\res[L,X] )]	\nonumber 
\end{align}
define two compatible Hamiltonian operators. Equivalently, the brackets $\{,\}_1$ and $\{,\}_2$ defined by
  \begin{equation}
\{f,g\}_i = <df, P_i dg>
\end{equation}
are two compatible Poisson brackets. Their explicit form is 
\begin{align} \label{e7.1}
    \{ u_n (x) , u_m(y) \}_1 &=  c_{nm} \big(  u_{n+m}(x) \de(x-y+n \ep) - u_{n+m}(x-m \ep) \de(x-y-m\ep)  \big) \\
    \{ u_n (x) , u_m (y) \}_2 &=  \sum_{l<m} [ u_{n+m-l}(x) u_l(x+(n-l)\ep) \de(x-y + (n-l)\ep) \label{e7.1b} \\
    &-u_l(x) u_{n+m-l} (x+(l-m)\ep) \de(x-y+(l-m)\ep)]  \nonumber \\
    &+ u_n(x) \left( \frac{(\L^n-\L^N)(\L^{-m}-1)}{1-\L^N} u_m(x) \de(x-y) \right)  \nonumber
\end{align}
where the constants $c_{nm}$ are defined by
\begin{equation} \label{e7.1b1}
c_{nm}=
\begin{cases}
1 &n>0,m>0 \\
-1 &n\leq0, m\leq0 \\
0 &\text{otherwise,}
\end{cases} 
\end{equation}
and the variables $u_n$ on the RHS are assumed to be zero if $n<-M$ or $n>N$ and $u_N =1$.
\end{prop}
\pf
The fact that $\{,\}_1$ and $\{,\}_2$ are two compatible Poisson brackets follows from the $R$-matrix theory developed in \cite{oe-ra89}and \cite{li-pa89}. This method has been already applied to the Toda lattice in \cite{bl-ma94}and to the two-dimensional Toda hierarchy in \cite{ca05}. Since this construction is well known we will give here only the basic steps.

Consider the associative algebra $\cGh$ of difference operators with arbitrary upper-bounded order
\begin{equation}
\label{ }
\cGh = \{ L = \sum_{k<\infty } u_k \L^k  \}
\end{equation}
which we identify with its dual $\cGh^*$ using the non-degenerate invariant inner product \eq{rr3}.
The map $R: \cGh \rightarrow \cGh$ given by $R(X) = X_+ - X_-$ and its skew-symmetric part $A= \frac12 (R-R^*)$ satisfy the modified Yang-Baxter equation
\begin{equation}
\label{ }
[R(X), R(Y) ] -R([R(X),Y]+[X,R(Y)])=-[X,Y] \quad 
\text{for every } X, Y \in \cGh
\end{equation}
thus, as a direct consequence of Lemma 1 of \cite{oe-ra89}, there are three compatible Poisson brackets on the algebra $\cGh$. The first Hamiltonian operator is simply given by \eq{e7}and the second one by the first line in \eq{e7b}. 

The next step is to perform a Dirac reduction to the affine subspace $\cG \subset \cGh$ of operators of the form \eq{a1}. While for $P_1$ the reduction is trivial, this is not the case for $P_2$ and a correction term is required. By taking this into account we obtain \eq{e7} and \eq{e7b}.

The third Poisson structure on the algebra $\cGh$ does not behave nicely under this reduction, as it has already been remarked in \cite{bl-ma94}, and it wasn't possible to find a close expression for the correction term, hence it will not be considered here.

Finally the compatibility of $\{,\}_1$ and $\{,\}_2$ after the reduction follows from the simple observation that 
\begin{equation}
\{,\}_2 \rightarrow \{,\}_2 + \lambda \{,\}_1 .
\end{equation}
under $u_0 \rightarrow u_0 + \lambda$.
\epf

\begin{rem}
As one can see from \eq{e7.1b}, when $N>1$ the second Poisson bracket contains a non-local term which can be rewritten as
\begin{equation}
\label{e7.4}
 u_n(x) \left( \frac{\L^n + \cdots + \L^{N-1}}{1+ \cdots + \L^{N-1}}(\L^{-m}-1) u_m(x) \de(x-y) \right) .
\end{equation}
This type of nonlocality is anyway well-defined when we regard the Poisson bracket as a formal power series in $\ep$ with coefficients that are differential polynomials. Expanding \eq{e7.4} we get indeed
\begin{equation}
\label{ }
 u_n(x) \sum_{k>0}c_k(n,m) \ep^k \frac{d^k}{dx^k}\left( u_m(x) \de(x-y) \right)
\end{equation}
where the generating function for the coefficients $c_k(n,m)$ is given by:
\begin{equation}
\label{ }
\frac{e^{n\la}+\cdots + e^{(N-1)\la}}{1+ \cdots + e^{(N-1)\la}}(e^{-m \la}-1) = \sum_{k>0} c_k(n,m) \la^k .
\end{equation}
\end{rem}

%%%%%%%%%%%%%%%%%%%%%%%%%%
\begin{rem}
Formulas \eq{e7}-\eq{e7b} already appeared in the literature. In particular for $N=1$, $M=1$ they reproduce the well-known Poisson pencil for the usual Toda lattice hierarchy \cite{ku85}, while the cases $N=2$, $M=1$ and $N=1$, $M=2$ where obtained in \cite{bl-ma94}. In the context of $q$-deformed $\mathcal{W}$-algebras they appeared in \cite{fr-re96, fr96} in the case $M=0$ (although with slight difference in the form of  $\{,\}_1$) and in \cite{ma-se95} for general $N$ ,$M$.

For convenience we report some examples below. Notice that in the $N\geq 2$ case the second Hamiltonian operator contains a non-local term of the form $(1+\L)^{-1}$. The entry $(P_i)^{nm}$ in the following matrices is an operator defined by
$P_i (\L^{-m} f) = \sum_{n=-M}^{N-1} ((P_i)^{nm} f )\L^n$. 
\begin{itemize}
\item $N=1$, $M=1$
\begin{equation} \label{ee1}
P_1= 
\begin{pmatrix}
0
& u_{-1} (1-\L^{-1})\\
(\L -1)u_{-1} & 0 
\end{pmatrix}
\end{equation}
\begin{equation} \label{ee2}
P_2=  
\begin{pmatrix}
u_{-1} (\L -\L^{-1}) u_{-1}
& u_{-1} (1-\L^{-1}) u_0 \\
u_0 (\L-1) u_{-1} 
& \L u_{-1} - u_{-1} \L^{-1} 
\end{pmatrix}
\end{equation}
\item $N=2$, $M=1$
\begin{equation} \label{ee3}
P_1= 
\begin{pmatrix}
0 
& u_{-1} (1-\L^{-1})
& 0\\
(\L-1) u_{-1}
& 0
& 0\\
0
& 0
& (\L-\L^{-1})  
\end{pmatrix}
\end{equation}
\begin{equation} \label{ee4}
P_2= 
\begin{pmatrix}
u_{-1} ( \L -1 - \L^{-1} + 2 \sig ) u_{-1}
& u_{-1} (1-\L^{-1})u_0 
& u_{-1}(2\sig - \L^{-1})u_1 \\
u_0 (\L-1)u_{-1}
& u_1 \L u_{-1} -u_{-1}\L^{-1}u_1
& \L u_{-1} - u_{-1} \L^{-2} \\
u_1 (\L-2 +2 \sig ) u_{-1}
& \L^2 u_{-1} - u_{-1} \L^{-1}
& \L u_0 - u_0 \L^{-1} - u_1^2 +2 u_1 \sig u_1  \\ 
\end{pmatrix}
\end{equation}
\item $N=1$, $M=2$
\begin{equation} \label{ee5}
P_1= 
\begin{pmatrix}
0 
& 0
& u_{-2} (1- \L^{-2})\\
0
& \L u_{-2} -u_{-2} \L^{-1}
& u_{-1} (1-\L^{-1})\\
(\L^2 -1) u_{-2}
& (\L -1)u_{-1}
& 0
\end{pmatrix}
\end{equation}
{\Small
\begin{equation} \label{ee6}
P_2= 
\begin{pmatrix}
u_{-2} (\L^2 +\L -\L^{-1}-\L^{-2}) u_{-2} 
& u_{-2} (\L +1 -\L^{-1} -\L^{-2} ) u_{-1} 
&  u_{-2} (1-\L^{-2}) u_0 \\
u_{-1} (\L^2 +\L -\L^{-1} -1) u_{-2}  
& u_0 \L u_{-2} -u_{-2} \L^{-1} u_0 +u_{-1}(\L -\L^{-1}) u_{-1}
& \L u_{-2} -u_{-2} \L^{-2} + u_{-1} ( 1- \L^{-1} ) u_0  \\
u_0 (\L^2 -1) u_{-2} 
& \L u_{-2} -u_{-2} \L^{-1} +u_0 (\L -1) u_{-1}
& \L u_{-1} -u_{-1} \L^{-1}
\end{pmatrix}
\end{equation}
}
\end{itemize}
\end{rem}

The flows of the bigraded Toda hierarchy can be expressed as Hamiltonian flows using the compatible Poisson brackets that we have just defined. 
The Hamiltonians are given by
\begin{equation} \label{bb1} 
  H_{\alpha, q} = \int  h_{\alpha, q} \ dx 
\end{equation}
and the Hamiltonian densities $h_{\alpha, q}$ by
\begin{align} \label{b28}
  &h_{\alpha, q} =  \frac{\Gamma ( 2- \frac{\alpha}N )}{\Gamma (q+3 - \frac{\alpha}N )} \res ( L^{q+2-\frac{\alpha}N} ) \qquad 
  \alpha = N-1, \dots , 0 \\
  &h_{\alpha, q} =  \frac{\Gamma ( 2+ \frac{\alpha}M )}{\Gamma (q+3 + \frac{\alpha}M )} \res ( L^{q+2+\frac{\alpha}M} ) \qquad 
  \alpha = 0, \dots , -M+1 \label{b28.1} \\
  &h_{-M,q} = 2 \frac1{(q+1)!} \res \left( L^{q+1} ( \log L - \frac12(\frac1M + \frac1N) c_{q+1} ) \right).  \label{b28.2}
\end{align}
Recall that $c_q = 1+ \dots +\frac1q$ and $c_0=0$.

The Hamiltonian densities are clearly elements of $\hat{\cA}$ and is simple to see that
\begin{prop}
  The grading of the Hamiltonian densities is 
  \begin{equation}
    \label{b29}
    \deg h_{\alpha, q } = q+2 + \mu_{\alpha} .
  \end{equation}
  where the $\mu_{\alpha}$'s are defined in \eq{b23.1}.
\end{prop}

The Hamiltonian densities can also be written as 
\begin{equation}
\label{d3}
h_{\alpha, p}=  \res B_{\alpha, p+1}
\end{equation}
where the operators $B_{\alpha, p}$ have been defined in \eq{d1}. We want to show that
\begin{prop}
\begin{equation}
  \label{e8}
  d H_{\alpha , p} =  B_{\alpha , p}.
\end{equation}
\end{prop}
\pf
Consider first the case $\alpha \geq 0$. Let's assume $L(t)$ depends on $t$ with $L(0)=L$ and denote by $X_t$ the derivative of $X$ w.r.t. $t$ evaluated at $t=0$.
One has 
\begin{equation}
\label{ }
( \tr (L^{\frac{p}N}))_t = p \tr(L^{\frac{p-1}{N}} (L^{\frac1N} )_t ) .
\end{equation}
Deriving the identity $(L^{\frac1N})^N$, multiplying by $L^{\frac{-N+p}N}$ and taking the trace we get
\begin{equation}
\label{ }
\tr (L^{\frac{p}N -1} L_t ) = N \tr(L^{\frac{p-1}N} (L^{\frac1N})_t).
\end{equation}
Comparing the last two equations and using definition \eq{rr5.1} we obtain
\begin{equation}
\label{ }
d \tr(L^{\frac{p}N}) = \frac{p}N L^{\frac{p}N-1}
\end{equation}
which gives \eq{e8} for $\alpha \geq 0$. The case $\alpha \leq 0$, $\alpha \not= -M$ follows from 
\begin{equation}
\label{ }
d \tr(L^{\frac{p}M}) = \frac{p}M L^{\frac{p}M-1}
\end{equation}
which is obtained in a similar way.

The case $\alpha = -M$ follows easily from the identity
\begin{equation}
\label{for5}
\tr (L^p (\log L)_t) = \left( \frac1{2N}+\frac1{2M} \right)\tr (L^{p-1} L_t).
\end{equation}

To prove this derive \eq{for1} with respect to $t$, then multiply on the left by $L^{p-1}$:
\begin{equation}
\label{ }
L^{p-1} L_t  = \sum_{k\geq 0} \frac1{k!} L^{p-1} ((\log_+ L)^k)_t e^{-N\ep\pa} \L^N .
\end{equation}
Both sides are difference operators, so it makes sense to take the trace. Using the fact that the trace 
\begin{equation}
\label{ }
\tr [ \log_+ L, A ] = \tr( N \ep (A)_x +2 N [W_- ,A] )
\end{equation}
is zero for any difference operator $A$, one obtains 
\begin{equation}
\label{ }
\tr (L^{p-1} L_t) = \tr(L^p (\log_+ L)_t) . 
\end{equation}
In a similar way one can show that
\begin{equation}
\label{ }
\tr (L^{p-1} L_t) = \tr(L^p (\log_- L)_t). 
\end{equation}
Combining the last two equations one obtains \eq{for5}
\epf

We want to show that the Hamiltonians $H_{\alpha,p}$ are in involution. This follows from a more general observation:
\begin{prop}
Let $f$, $g$ be two functions of $L$ such that $[ df, L] = 0$ and $[dg , L]=0$. Then they are in involution with respect to $\{,\}_1$ and $\{,\}_2$. Moreover $L_t = [(df)_+, L] = [-(df)_-, L]$ is a well-defined flow on $\mathcal{G}$.
\end{prop}
\pf
The involution property of $f$ and $g$ simply follows from: 
\begin{align}
\label{ }
\{f,g\}_1 &= <df, P_1 (dg)> =  <df, [(dg)_+ ,L]> =<[L,df],(dg)_+>=0 \\
\{f,g\}_2 &= <df,P_2(dg)> = <df, [L, (L dg)_-]> =<[df,L],(L dg)_-> =0
\end{align}
where we have used the invariance property of the inner product: 
\begin{equation}
\label{ }
<[L,X],Y> = <X,[Y,L]> .
\end{equation}
The fact that the flow 
\begin{equation}
\label{ }
L_t = [(df)_+, L] = [-(df)_-, L]
\end{equation}
is well-defined simply follows from the observation that the order in $\L$ of the second term is $\geq -M$ while the order of the third term is $\leq N-1$, hence $L_t$ is a well-defined vector field on $\cG$.  
\epf

The bi-Hamiltonian formulation of the bigraded Toda hierarchy is then given by the following
\begin{thm}
The flows of the bigraded Toda hierarchy are Hamiltonian with respect to the Poisson brackets \eq{e7.1}-\eq{e7.1b} and the Hamiltonians given in \eq{bb1}-\eq{b28.2} are in involution with respect to both Poisson brackets. Indeed the flows \eq{b5} can be written as 
\begin{equation} \label{b29.1}
\ep \frac{d}{d t^{\alpha, q }} u_i(x)  = \{ u_i(x) , H_{\alpha, q} \}_1  
\end{equation} 
for $q \geq 0$ and $-M \leq \alpha \leq N-1$, and we have the recursion relations
\begin{equation}
\label{e11}
\{ \cdot , H_{\alpha,p} \}_2 = (p+2+\mu_{\alpha} ) \{ \cdot , H_{\alpha, p+1} \}_1 +  \sum_{\beta=-M}^{N-1} R_{\alpha}^{\beta} \{ \cdot , H_{\beta, p} \}_1
\end{equation}
for $p \geq -1$, where
\begin{equation}
\label{e10}
R_{\alpha}^{\beta} = (\frac1N+\frac1M) \delta_{\alpha, -M} \delta^{\beta}_0
\end{equation}
with  $-M \leq \alpha, \beta  \leq N-1$ and 
\begin{equation}
\label{ }
\mu_{\alpha} = \begin{cases}
      & -\frac{\alpha}N \quad \alpha \geq 0 , \\
      & \frac{\alpha}M \quad \alpha \leq 0.
\end{cases}
\end{equation}
\end{thm}
\pf 
The involution property of the Hamiltonians follows from the previous Proposition and from the fact that $[d H_{\alpha,p} , L ]=0$.

Equation \eq{b29.1} simply follows from \eq{e7} and \eq{e8} that give
\begin{equation}
\label{ }
P_1(dH_{\alpha, p}) = [ (B_{\alpha,p})_+ , L ]  = [- (B_{\alpha,p})_- , L ] ;
\end{equation}
these are exactly the Lax equations of the bigraded Toda hierarchy as defined in \eq{b5}-\eq{b6}.

From the definition \eq{d1} we easily get
\begin{equation}
  \label{e9}
  L B_{\alpha, p} = (p+2+\mu_{\alpha} ) B_{\alpha,p+1} +R_{\alpha}^{\beta} B_{\beta,p}.
\end{equation}
Applying $P_1$ on both sides and observing that
\begin{equation}
\label{ }
P_1(L X) =  P_2(X) \quad \text{when} \quad [L, X]=0
\end{equation}
we obtain the recursion relation \eq{e11}.
\epf

%%%%%%%%%%%%%%%%%%%%%%
\section{Tau function}
In this section we will prove the existence of a tau-function for the bigraded extended Toda hierarchy, following the approach of \cite{dkjm83} for the KP hierarchy.

While the Hamiltonians $H_{\alpha,p}$ are obviously conserved by the flows of the hierarchy, the Hamiltonian densities have interesting derivatives with respect to $t^{\beta,q}$ which are given by the following 
\begin{lemma} \label{lem11}
  The following formula holds
  \begin{equation}
    \label{b35}
    \ep \frac{\pa h_{\alpha, p-1}}{\pa t^{\beta ,q} } = 
    \begin{cases}
       \res [ -(B_{\beta, q})_- , B_{\alpha,p} ] 
       &\alpha =0 \dots N-1 \\
       \res [ (B_{\beta, q})_+, B_{\alpha,p}] & \alpha =0  \dots -M+1 \\
      \frac2{p!} \res \left( [ -(B_{\beta, q})_- , L^p (\frac1{2N}\log_+ L - \frac{\kappa_p}2 )  ] + [ (B_{\beta, q})_+, L^p (\frac1{2M}\log_- L - \frac{\kappa_p}2 ) ] \right)
 & \alpha =-M  .
    \end{cases}
  \end{equation}
for $p,q \geq 0$ and $\kappa_p=\frac12 (\frac1M + \frac1N) c_p$.
\end{lemma}

Notice that in the last line of the previous formula we are apparently performing the undefined operation of taking the residue of a differential-difference operator (since it contains $\log_+ L$ and $\log_- L$. However the derivative operators cancel in the RHS as one can see from the following equivalent formula
{\Small
\begin{equation}
  \label{b35.1}
  \ep \frac{\pa h_{-M ,p-1} }{\pa t^{\beta, q}} = \frac2{p!} \res \left( \frac{\ep}2 L^p (B_{\beta,q} )_x  + 
[ -(B_{\beta,q} )_- , L^p ( (\log L)_- -\frac12 \kappa_p ) ] + 
[ (B_{\beta,q} )_+ , L^p ( (\log L)_+ -\frac12 \kappa_p ) ] \right)
\end{equation}
}
\pf
The proof is easily obtained by using Lemma \ref{lemD} and \eq{d3}.
\epf

It is easy to see that each differential polynomial in \eq{b35} is homogeneous in the grading \eq{b8}, with degree given by
\begin{equation}
  \label{b38}
  \deg ( \frac{\pa h_{\alpha , p-1}}{\pa t^{\beta, q}}) = q+p+2 + \mu_{\alpha} + \mu_{\beta} .
\end{equation}

%We define the differential polynomials $\Omega_{\alpha, p; \beta, q}$ 
\begin{defi}
The differential polynomials $\Omega_{\alpha, p; \beta, q} \in \hat{\cA}$ are uniquely defined by 
\begin{equation}
  \label{b40}
  \frac1{\ep} (\L -1) \Omega_{\alpha, p; \beta ,q} := \frac{\pa h_{\alpha, p-1}}{\pa t^{\beta,q}} ,
\end{equation}
with the requirement that they are homogeneous of degree
\begin{equation}
  \label{c1}
  \deg \Omega_{\alpha,p;\beta,q} = q+p+2 + \mu_{\alpha} + \mu_{\beta}
\end{equation}
for $p,q \geq 0$, $\alpha, \beta = N-1, \dots , -M$ 
and by 
\begin{equation}
  \label{c3}
  \Omega_{-M,0;-M,0} = \ep^2 (\L -1)^{-1} (1- \L^{-M} )^{-1} (\log u_{-M} )_{xx} .
\end{equation}
\end{defi}
\pf
It is clear that the differential polynomials \eq{b35} are always in the image of $\L -1$, hence $\Omega_{\alpha,p;\beta,q}$ exists in $\hat{\cA}$. The ambiguity given by the kernel of $\L -1$ is fixed by the requirement of homogeneity, as far as $\deg \Omega_{\alpha,p;\beta,q} \not= 0$. This only happens when $p=q=0$ and $\alpha=\beta=-M$ in which case we require \eq{c3}. 
\epf

We now state the important property of tau--symmetry that holds due to the particular normalization of the flows chosen above. 
\begin{thm}
  The Hamiltonian densities $h_{\alpha, p}$ satisfy the tau symmetry , i.e. the following identities hold
  \begin{equation}
    \label{b30}
    \frac{\pa h_{\alpha, p-1}}{\pa t^{\beta, q}} = \frac{\pa h_{\beta, q-1}}{\pa t^{\alpha, p}}
  \end{equation}
  for $\alpha, \beta =N-1, \dots , -M$ and $p,q \geq 0$.
\end{thm}
\pf
One essentially uses Lemma \ref{lem11} and shows that for every choice of indices on the LHS it is possible to rewrite the RHS in such a way that \eq{b30} holds. 

For example consider first the simple case $\alpha \geq 0$ and $\beta \geq 0$. Then we have
\begin{equation}
  \label{d8}
  \ep \frac{\pa h_{\alpha, p-1}}{\pa t^{\beta,q}} =  \res [ -(B_{\beta,q})_- , B_{\alpha,p} ] .
\end{equation}
Using the fact that $\res [A_- ,B_-] = \res [A_+ ,B_+] = 0$ for every pair of difference operators $A$, $B$, we rewrite this as 
\begin{equation}
  \label{d9}
   \res [  ( B_{\alpha,p})_+ ,  B_{\beta,q} ] .
\end{equation}
Then, using the fact that $[  B_{\alpha,p},  B_{\beta,q} ] = 0$, we have that this is equal to
\begin{equation}
  \label{d10}
   [ -(   B_{\alpha,p} )_- ,  B_{\beta,q}  ] = \ep \frac{\pa h_{\beta, q-1}}{\pa t^{\alpha, p}} 
\end{equation}
so the formula is proved in this case.

Most other cases are proved in a similar way. We limit ourselves to the  case $\alpha= \beta =- M$ that is the more complicated.
It is easy to see that 
\begin{subequations} \label{e3}
\begin{align}
  \ep \frac{\pa h_{-M, p-1}}{\pa t^{-M,q}} &= 4 \frac1{p! q!} \res
  \left([ -(L^q \log L)_- , L^p \frac1{2N} \log_+ L ] + [ (L^q \log L)_+ , L^p \frac1{2M} \log_- L ] \right) \\
  &+\ep \kappa_p \frac{\pa h_{-M, q-1}}{\pa t^{0, p-1}} -\ep \kappa_q \frac{\pa h_{-M, p-1}}{\pa t^{0, q-1}} 
      +\ep \kappa_p \kappa_q \frac{\pa h_{0,p-2}}{\pa t^{0,q-1}} .
\end{align}
\end{subequations}
Since the second line in this formula is symmetric in $p$, $q$, we have just to show that the residue on the RHS is symmetric in $p$, $q$.

Using the definitions \eq{a8}- \eq{a9} we have that such residue is written as
\begin{subequations}   \label{e4}
\begin{align}
&\res ( [ -(L^q \log L)_- , L^p \frac1{2N} \log_+ L ] + [ (L^q \log L)_+ , L^p \frac1{2M} \log_- L ] ) = \\
  &=\res [ -(L^q \log L)_- , L^p (\log L)_- ] + \res [ (L^q \log L)_+, L^p (\log L)_+ ] +\\
  &+\res [L^p \frac{\ep}2 \pa ,L^q \log L ] .
\end{align}
\end{subequations}

Using as above the fact that $\res [A_- ,B_-] = \res [A_+ ,B_+] = 0$ for every difference operators $A$, $B$, one rewrites (\ref{e4}b) as
\begin{subequations}  \label{e5}
\begin{align}
  &\res [ -(L^q \log L)_- , L^p (\log L)_- ] + \res [ (L^q \log L)_+, L^p (\log L)_+ ] = \\
  &=\res [ (L^p( \log L)_-)_+ , L^q (\log L)_+ ] + \res [ (L^p( \log L)_-)_+, L^q (\log L)_- ] +\\
  &- \res [ (L^p( \log L)_+)_- , L^q (\log L)_+ ] - \res [ (L^p( \log L)_+)_-, L^q (\log L)_- ],
\end{align}
\end{subequations}
and, using the fact that $[ \log_+ L , \log_+ L ] =0 = [ \log_- L, \log_- L ]$, the term (\ref{e4}c) becomes
\begin{subequations}  \label{e6}
\begin{align}
  & \res [L^p \frac{\ep}2 \pa ,L^q \log L ] = \res [L^q \log L, L^q (-\frac{\ep}2 \pa) ] + \\
  &+ \res [ L^p (\log L)_+, L^q (\log L)_+ ] - \res [ L^p (\log L)_-, L^q (\log L)_- ] .
 \end{align}
\end{subequations}
Combining equations \eq{e5} and \eq{e6} one easily obtains the desired result.
\epf

Using the terminology of \cite{du-za01}, we have that this tau structure is compatible with spatial translations , i.e. the Hamiltonian $\bar{h}_{-M,0}$ corresponds to the $x$-translations
  \begin{equation}
    \label{b34}
    \frac{\pa}{\pa t^{-M, 0} } \cdot =  \frac{\pa}{\pa x} \cdot
  \end{equation}
since $A_{-M, 0} = \ep \frac{\pa}{\pa x}$.

The property of tau symmetry implies that $\Omega_{\alpha, p; \beta, q}$ is symmetric under the exchange of the pair of indices $(\alpha,p)$ and $(\beta, q)$. Form the definition \eq{b40} together with \eq{b30} it follows moreover that $\frac{\pa \Omega_{\alpha, p; \beta, q}}{\pa t^{\sigma ,k}}$ is symmetric with respect to all three pairs of indices. From these properties we obtain the existence of a tau-function for the hierarchy. We have indeed the following:
\begin{thm}
  For any solution $u_k$, $k=N-1, \dots, -M$, of the extended bigraded Toda hierarchy there exists a function $\tau = \tau ( x, {\mathbf t}, \ep )$, depending on the spatial variable $x$, on all the time variables ${\mathbf t} = \{ t^{\alpha, q}, -M \leq \alpha \leq N-1, q \geq 0 \}$ and on $\ep$, such that 
  \begin{equation}
    \label{b31}
    \Omega_{\alpha, p; \beta, q} = \ep^2 \frac{\pa^2 \log \tau}{\pa t^{\alpha,p} \pa t^{\beta, q}} . 
  \end{equation}
\end{thm}

Moreover, if we require that 
\begin{equation}
  \label{b37}
  \frac{\pa \log \tau}{\pa t^{-M,0}}  =  \frac{\pa \log \tau}{\pa x} 
\end{equation}
then the following Corollary holds
\begin{corol}
  The densities of the Hamiltonians of the extended bigraded Toda hierarchy can be expressed in terms of the tau function by the following formula
  \begin{equation}
    \label{b36}
    h_{\alpha, p} = \ep (\L -1) \frac{\pa \log \tau}{\pa t^{\alpha, p+1}}
  \end{equation}
for $p \geq -1$ and $\alpha = N-1, \dots ,-M$.
\end{corol}
\pf By definition \eq{b40} we have
\begin{equation}
  \label{e1}
  \ep \frac{\pa h_{\alpha, p}}{\pa x} = (\L -1 ) \Omega_{\alpha, p+1; -M,0}.
\end{equation}
Since both sides in the previous equation are uniquely determined homogeneous elements of $\hat{\cA}$ we can integrate in $x$ without ambiguity and obtain
\begin{subequations} \label{e2}
  \begin{align}
    h_{\alpha,p} &=  \sum_{k=1}^{\infty} \frac{(\ep \pa_x)^{k-1}}{k!} \Omega_{\alpha,p;-M,0} \\
    &= \ep \sum_{k=1}^{\infty} \frac{(\ep \pa_x)^k}{k!} \frac{\pa \log \tau}{\pa t^{\alpha,p+1}} \\
    &= \ep (\L -1) \frac{\pa \log \tau}{\pa t^{\alpha, p+1}}.
  \end{align}
\end{subequations}
\epf

From \eq{b31}, \eq{b36} we see that the coordinates $u_k$ are not normal coordinates in the sense of \cite{du-za01}.

%%%%%%%%%%%%%%%%%%%%%%%%%%%%%%%

\section{The dispersionless limit and Frobenius manifolds}
In this section we will consider the dispersionless limit of the extended bigraded Toda hierarchy and provide generating functions for the Poisson brackets and the relative flat metrics. We show that the Frobenius manifolds associated to these hierarchies  are those given on the orbit spaces of extended affine Weyl groups of the $A$-series. 

For convenience we consider a slightly different normalization of the hierarchy: we multiply the first Poisson structure \eq{e7.1} and all the times of the hierarchy by $(-1)^N$.

The dispersionless Poisson brackets $\{ , \}_i^{disp}$ are obtained as the leading term of the dispersive brackets \eq{e7.1} and \eq{e7.1b} in the $\ep \rightarrow 0$ limit, i.e.
\begin{equation}
\label{}
\{ u_n(x), u_m(y) \}_i = \ep \{u_n(x), u_m(y) \}_i^{disp} +O(\ep^2) .
\end{equation}

The explicit form of the dispersionless Poisson brackets is
\begin{align}
\label{disp1}
\{ u_n(x) , u_m(y) \}_1^{disp} &= (-1)^N  c_{nm} \left[ (n+m) u_{n+m} (x) \de'(x-y) +m u'_{n+m} (x) \de(x-y) \right] , \\
\{ u_n(x) ,u_m(y) \}_2^{disp} &=  \sum_{l<m} \big[ (n+m-2l) u_l u_{n+m-l} \de'(x-y) \label{disp1.1}\\
			&+(n-l) u'_l u_{n+m-l} \de(x-y) +(m-l) u_l u'_{n+m-l} \de(x-y) \big] \notag \\
			&+ \frac{m}N (n-N) u_n u_m \de'(x-y) + \frac{m}N (n-N) u_n u'_m \de(x-y) . \notag
\end{align}
where the constants $c_{nm}$ are defined in \eq{e7.1b1} and the variables $u_n$ on the RHS are assumed to be zero if $n<-M$ or $n>N$ and $u_N =1$.

The dispersionless Poisson brackets \eq{disp1}, \eq{disp1.1} are of hydrodynamic type, i.e. they are of the form
\begin{equation}
\label{ }
\{ u_n(x),u_m(y) \} = g^{nm} \de'(x-y) + \Gamma_k^{nm} u'_k \de(x-y) .
\end{equation}
It is a well-known result \cite{du-no89} that such brackets satisfy the Jacobi identity  if and only if the coefficients $g^{nm}$  define a flat contravariant metric and the coefficients $\Gamma_{ij}^k = - g_{il} \Gamma_j^{lk}$ are the Christoffel symbols of the Levi-Civita connection associated with $g_{ij}$.

We want to obtain generating functions for the metric $g^{nm}_{(i)}$ associated to the dispersionless Poisson bracket $\{ ,\}_i^{disp}$. We will first write generating functions for the dispersionless Poisson brackets.

We can organize the dependent variables in a dispersionless Lax function
\begin{equation}
\label{ }
\la (p,x) = p^N + u_{N-1} p^{N-1} + \dots + u_{-M} p^{-M}
\end{equation}
and define the generating functions by
\begin{equation}
\label{ }
\{ \la(p,x) , \la(q,y) \}_i^{disp} =  \sum_{nm} \{ u_n(x) , u_m(y) \}_i^{disp} p^n q^m .
\end{equation}
\begin{prop}
The generating functions for the first and second dispersionless Poisson brackets are
\begin{multline}
\{ \la(p,x) , \la(q,y) \}_1^{disp} = (-1)^N \left[ \frac1{p^{-1} -q^{-1} } \left( \frac{\pa}{\pa q} \la(q,y) -\frac{\pa}{\pa p} \la(p,x) \right) \de'(x-y) \right.\\
\left. + \frac{p^{-1} q^{-1} }{(p^{-1} -q^{-1})^2} \left( \la(q,x) + \la(p,y) -\la(p,x) -\la(q,y) \right) \de'(x-y) \right]
\end{multline}
\begin{multline}
\{ \la(p,x) ,\la(q,y) \}_2^{disp} = \frac1N pq \frac{\pa}{\pa p}\la(p,x) \frac{\pa}{\pa q} \la(q,y) \de'(x-y) + \\
+ \frac{1}{p^{-1} -q^{-1}} \left( \la(p,x) \frac{\pa}{\pa q} \la(q,y) - \la(q,y) \frac{\pa}{\pa p} \la(p,x) \right) \de'(x-y)  \\
+ \frac{p^{-1} q^{-1}}{(p^{-1} -q^{-1})^2} \left( \la(p,y) \la(q,x) - \la(p,x)\la(q,y) \right) \de'(x-y)
\end{multline}
\end{prop}

If we define the generating functions for the contravariant metrics $g_{(i)}^{nm}$ to be
\begin{equation}
\label{ }
( d\la(p), d\la(q) )_i = \sum_{nm} g_{(i)}^{nm} p^n q^m . 
\end{equation}
we have
\begin{prop}
The generating functions for the contravariant metrics associated to the dispersionless Poisson brackets \eq{disp1}, \eq{disp1.1} are
\begin{align}
\label{lekker139}
   (d \la(p) , d \la(q) )_1 &= (-1)^N \frac{\la'(q) - \la'(p)}{p^{-1}-q^{-1}} ,  \\
   (d \la (p) , d \la(q) )_2  &= \frac1N pq \la'(p) \la'(q) + \frac1{p^{-1} - q^{-1}} (   \la(p) \la'(q) -\la(q) \la'(p) ). \label{lekker139.1}
\end{align}
\end{prop}

Now we want to show that this tau-symmetric bihamiltonian structure defines a Frobenius manifold, for fixed $N$, $M$, that coincides with the one defined on the orbit space of the extended affine Weyl group $\tilde{W}^{(N)}(A_{N+M-1})$.
For an introduction to and the main definitions of Frobenius manifolds see \cite{du96}.

Frobenius manifolds were constructed on the orbit spaces of extended affine Weyl groups in \cite{du-za98}. In particular a Frobenius manifold $M_{\tilde{W}^{(k)}(R)}$ is associated to each pair $(R,k)$ where $R$ is an irreducible reduced root system and $k$ is a choice of simple root, in a range that depends on $R$. In the case where the root system is $A_l$ one can choose a root $k$ with $1 \leq k \leq l$. We will consider here the case of $A_{N+M-1}$ with the choice of the $N$-th simple root. 

In this case an explicit representation of the Frobenius manifold $M_{\tilde{W}^{(N)}(A_{N+M-1})}$ has been given in \cite{du-za98} as the space of affine trigonometric polynomials of bidegree $(N,M)$. We can rewrite this result in the following form

\begin{prop}[\cite{du-za98}]
Consider the space $M_{N,M}$ of rational functions of  $z$ of the form
\begin{equation}
\label{ }
\la (z) = z^N + u_{N-1} z^{N-1} + \dots + u_{-M} z^{-M}
\end{equation}
with $u_{-M} \not= 0$. The invariant inner product of two vectors $\pa'$ and $\pa''$ tangent to $M_{N,M}$ at a point $\la$ is given by
\begin{equation}
\label{lekker182}
<\pa',\pa''> = (-1)^N \sum_{|\la | < \infty} \res_{d\la =0} \frac{\pa' (\la \ dz) \pa''(\la \ dz)}{z^2 d\la} 
\end{equation}
while the intersection form is given by
\begin{equation}
\label{lekker183}
(\pa', \pa'') = \sum_{|\la| < \infty} \res_{d\la=0} \frac{\pa'(\log \la \ dz) \pa''(\log \la \ dz)}{z^2 d \log \la} .
\end{equation}
In these expressions one sums the residues at the points where $d\la=0$ and $\la$ doesn't have a pole.
The Euler vector field $E$ and the unit vector field $e$ are given by
\begin{equation}
\label{ }
E= \sum_{k=-M}^{N-1} \frac{N-k}N u_k \frac{\pa}{\pa u_k} , \qquad e= (-1)^N \frac{\pa}{\pa u_0} .
\end{equation}
These definitions provide $M_{N,M}$ with a structure of Frobenius manifold and such Frobenius manifold is isomorphic to $M_{\tilde{W}^{(N)}(A_{N+M-1})}$.
\end{prop}
This is an instance of the general construction of Frobenius manifolds on the Hurwitz space of branched coverings over $\bar{\C}$. In particular $M_{N,M}$ is (a covering of) the Hurwitz space denoted by $\hat{M}_{0;N-1,M-1}$ in \cite{du96}.

The Frobenius manifolds provide the main invariant of tau-symmetric Poisson pencils in the classification theory of Dubrovin and Zhang \cite{du-za01}. Such invariant in particular depends only on the dispersionless limit of the bihamiltonian structure. 

We now show that such Frobenius manifold coincides with $M_{\tilde{W}^{(N)}(A_{N+M-1})}$.

\begin{prop}
The Frobenius manifold associated to the tau-symmetric Poisson structure of the bigraded extended Toda hierarchy is isomorphic to $M_{\tilde{W}^{(N)}(A_{N+M-1})}$.
\end{prop}
\pf
We know from \cite{du-za01} that a Frobenius structure is uniquely associated to every tau-symmetric Poisson structure. A direct  way to identify it is to make use of the one to one correspondence between Frobenius manifolds and quasihomogeneous flat pencils of metrics (see \cite{du98}). The flat pencil of metrics $g_{(1)}^{nm}$, $g_{(2)}^{nm}$ is indeed quasihomogeneous of degree $d=1$: if we define the function 
\begin{equation}
\label{ }
\tilde{\tau} = \frac1{M} \log u_{-M}
\end{equation}
we obtain,  following the prescription of \cite{du98}, the vector fields
\begin{equation}
\label{ }
e := \nabla_1 \tilde{\tau} = (-1)^N \frac{\pa}{\pa u_0} \qquad
E := \nabla_2 \tilde{\tau} = \sum_{k=-M}^{N-1} \frac{N-k}{N} u_k \frac{\pa}{\pa u_k}
\end{equation}
where $\nabla_i$ is the gradient associated to the metric $g_{(i)}^{nm}$. They satisfy the requirements for quasihomogeneity
\begin{align}
\label{}
    \cL_E g_{(2)}^{nm}     &= 0   \\
    \cL_e g_{(1)}^{nm} &=0 \\
    \cL_e g_{(2)}^{nm} &= g_{(1)}^{nm} 
\end{align}
and $[e,E]=e$, where $\cL$ is the Lie derivative.

Hence \cite{du98} we have that $e$ and $E$ are, respectively, the unit and the Euler vector fields of the associated Frobenius manifold. Since they coincide with those of $M_{M,N}$, we only need to check that the pencil of metrics is the same. 

Let's start from the first metric. We want to show that the contravariant metric $g_{(1)}^{ij}$ with generating function \eq{lekker139} is the inverse of the covariant metric given by \eq{lekker182}, i.e.
\begin{equation}
\label{huis1}
\sum_{m=-M}^{N-1} g_{(1)}^{nm} g_{mk} = \de_{nk}
\end{equation}
where from \eq{lekker182} we have
\begin{equation}
\label{ }
g_{mk} = <\frac{\pa}{\pa u_m},\frac{\pa}{\pa u_k}>=(-1)^N \sum_{|\la|<\infty} \res_{\la_z=0} \frac{z^{m+k-1}}{z \la_z} dz .
\end{equation}
If we multiply \eq{huis1} by $w^n$, sum over $-M \leq n \leq N-1$ and then use the generating function \eq{lekker139} we obtain that we have to prove
\begin{equation}
\label{huis2}
\sum_{|\la| < \infty} \res_{\la_z=0} \frac{(\la_w(w) - \la_z(z) ) z^{k-1}}{(z^{-1} -w^{-1}) z \la_z} dz = w^k .
\end{equation}
In the LHS the term $\la_z(z)$ in the numerator doesn't contribute, since it cancels with the denominator and gives a function without poles in $\la_z=0$. Hence the LHS is given by the following sum of three residues
\begin{equation}
\label{huis3}
(\res_{z=0} + \res_{z=\infty} + \res_{z=w} ) \frac{w \la_w(w) z^k}{(z-w) z \la_z} dz .
\end{equation}
The residue in $z=w$ gives exactly the desired result $w^k$, while it is easy to show that the other two residues are zero.

The fact that the second metric $g_{(2)}^{nm}$ is the inverse of the intersection form \eq{lekker183} is shown in the same way, using the generating function \eq{lekker139.1}.  One has to prove
\begin{multline}
\label{ }
\sum_{|\la| < \infty} \res_{\la_z=0} \frac{w \la_w(w) z^{k-1} }{N \la} dz - \sum_{|\la|< \infty} \res_{\la_z=0} \frac{\la(w) z^{k-1} }{(z^{-1}-w^{-1}) z \la} dz + \\
+ \sum_{|\la|<\infty} \res_{\la_z=0} \frac{\la_w(w) z^{k-1}}{(z^{-1} -w^{-1})z \la_z}dz = w^k .
\end{multline}
The first two terms on the LHS vanish since they don't have poles in $\la_z=0$. The third term gives exactly the same sum of residues \eq{huis3} as before.
\epf

\paragraph{\bf Acknowledgments}
This work was supported by the EPSRC grant GR/S48424/01 and by the EU GIFT project (NEST- Adventure Project no. 5006).
I would like to thank Prof. Boris Dubrovin for discussions related to this work.

%%%%%%%%%%%%%%%%%%%%%%%%%%%%%%%%%%%%%%%%%%%%%%%%%%%%%%%%%%%%%%%%%
\appendix

\end{document}